\documentclass[5p,twocolumn]{elsarticle}
\usepackage[utf8]{inputenc}
\usepackage[pdfencoding=auto]{hyperref}
\usepackage{multirow}
\usepackage{comment}
\usepackage{xspace}
\usepackage{balance}
\usepackage{multirow}
\usepackage[table,xcdraw]{xcolor}
\usepackage{mathtools}
\usepackage{booktabs}
\usepackage{amssymb}
\usepackage{pdfpages}
\usepackage{graphicx}
\usepackage{caption}
\usepackage{subcaption}
\usepackage{caption, subcaption}
\usepackage{amsthm}
\usepackage{float}
\usepackage{url}
\usepackage{cleveref}
\usepackage{flushend}
\usepackage{breakurl}

\makeatletter
\def\ps@first{%
   \let\@oddhead\@empty
   \let\@evenhead\@empty
   \def\@oddfoot{}
   \let\@evenfoot\@oddfoot
}

\newcommand{\TOOL}{DPMon\xspace}

\author[1]{Martino Trevisan\corref{cor1}}

\address[1]{University of Trieste, Italy}

\title{\TOOL: a Differentially-Private Query Engine for Passive Measurements}

\cortext[cor1]{Corresponding author: Martino Trevisan (martino.trevisan@polito.it)\\
\textit{Email address:} martino.trevisan@units.it
}

\begin{document}

\begin{abstract}

\noindent
Passive monitoring is a network measurement technique which analyzes the traffic carried by an operational network. It has several applications for traffic engineering, Quality of Experience monitoring and cyber security. However, it entails the processing of personal information, thus, threatening users' privacy.

\noindent
In this work, we propose \TOOL, a tool to run privacy-preserving queries to a dataset of passive network measurements. It exploits differential privacy to perturb the output of the query to preserve users' privacy. \TOOL can exploit big data infrastructures running Apache Spark and operate on different data formats. We show that \TOOL allows extracting meaningful insights from the data, while at the same time controlling the amount of disclosed information.

\end{abstract}

\maketitle

\begin{table*}[!h]
    \centering
    \small
    \begin{tabular}{c|l|l}
        \textbf{Nr} & \textbf{Code metadata description} & \textbf{Value} \\
        \hline
        C1 & Current code version  & 1.0 \\
        C2 & Permanent link to code/repository used for this code version  & \url{https://github.com/marty90/DPMon} \\
        C3 & Permanent link to reproducible capsule  & - \\
        C4 & Legal code license  & GNU General Public License v3.0 \\
        C5 & Code versioning system used  & \texttt{git} \\
        C6 & Software code languages, tools and services used  & Python\textsuperscript{TM} \\
        C7 & Compilation requirements, operating environments and dependencies  & Python\textsuperscript{TM}, optionally Apache Spark\textsuperscript{TM} \\
        C8 & If available, link to developer documentation/manual  & \url{https://marty90.github.io/DPMon} \\
        C9 & Support email for questions & \href{mailto:martino.trevisan@dia.units.it}{\texttt{martino.trevisan@dia.units.it}} \\
    \end{tabular}
    \caption{Code Metadata}
    \label{tab:metadata}
\end{table*}

\section{Introduction}
\label{sec:intro}

In recent years, the volume of data generated by ITC systems has significantly increased, creating tension between data and users’ privacy. Computer networks also accumulate vast amounts of data through network equipment such as firewalls, routers, and passive meters, providing rich datasets on network activity. These data are essential for monitoring users' Quality of Experience, detecting security incidents and analyzing human behavior. However, network data often contain sensitive personal information as it can reveal the identity and habits of users. Therefore, implementing appropriate measures is crucial. Unfortunately, conventional approaches to data anonymization, such as k-anonymity and its variants, are not well-suited for network contexts, posing challenges for sharing network data among researchers, companies, and system administrators.

Differential privacy (DP)~\cite{dwork2006calibrating} has emerged as a practical method for anonymizing data releases and queries on private datasets. Unlike k-anonymity, DP is not limited to tabular data and provides robust guarantees against post-processing attacks by adding controlled noise to the released data. This approach can be applied in networking, where the value of network data lies in its ability to address diverse and complex queries across various domains, including user behavior and network performance. Balancing data utility with individual privacy is crucial for any system handling personal information~\cite{li2009tradeoff}, and DP offers a means to achieve this balance.

This paper presents \TOOL, a novel open-source tool for executing differentially-private queries on network measurements. In developing this tool, we have identified suitable methodologies and technologies for applying DP to network data. Leveraging differentially private aggregate functions, \TOOL enables the execution of various queries on the data, encompassing volumetric and user-centric metrics. \TOOL is modular and supports data formats of various flow exporters, including NetFlow records. It can process data locally or harness the capabilities of a Spark big data cluster. We validate \TOOL in a real use case, demonstrating its ability to extract meaningful metrics from a dataset obtained from a campus network. It allows characterizing network usage and user habits, while, at the same time, offering strong guarantees of users' privacy.

We believe \TOOL offers several advantages for researchers, network practitioners and other stakeholders in the Internet community. For network practitioners, \TOOL facilitates QoE measurement, incident detection and traffic accounting. Within organizations, \TOOL enhances network management by providing cross-departmental visibility and promoting collaboration among technical units. Researchers can benefit from \TOOL by studying the impacts of new protocols and human behaviour, ultimately contributing to the development of faster and safer networks. Moreover, \TOOL, enables external actors to benefit from network data, as it allows private access to insights on network traffic. This may ultimately foster the data economy, enabling the sharing of business-valuable analytics while respecting users' privacy. For instance, public bodies or companies could conduct studies using Internet traffic without harming citizens' privacy. Overall, the privacy-aware monitoring capabilities of \TOOL foster cooperation and trust among network stakeholders, driving the advancement of more efficient and secure networks.
\section{Related Work}
\label{sec:related}

Network monitoring is a threat to user privacy~\cite{rfc7258} as it entails the processing of Personally Identifiable Information (PII), such as users' IP addresses. Moreover, other information contained in network measurements can disclose features of the monitored individuals that should not made public. For instance, the measurements at the DNS level can indirectly expose the browsing history of the individuals. The trend towards encryption~\cite{chan2018monitoring} mitigates this issue, but even encrypted protocols like TLS do disclose some information~\cite{khatouni2017privacy} such as the domain of the visited websites.

To allow privacy-preserving network monitoring, recent works have leveraged different techniques for data anonymization. The earliest proposal has been \emph{k}-anonymity, a technique that seeks to make an individual indistinguishable from other \emph{k-1} individuals in the released data~\cite{samarati_kanon_1998}. \emph{k}-anonymity is designed to work on tabular data, and it struggles in the case of high-dimensional datasets. Thus, it is hard to apply it to network measurements where the number of dimensions can be huge and event not known \emph{a priori}. A branch of works studying techniques to anonymize network data proposed approaches inspired from \emph{k}-anonymity so that it can fit the networking field~\cite{favale2021alpha,jha2023practical}.

Later on, differential privacy~\cite{dwork2006calibrating} was proposed as a different technique to release privacy-sensitive data. It consists of adding noise to the released data so that the contribution of a single individual is limited under a controllable threshold. Differently from the k-anonymity, it can be easily applied to the result of a query to a private dataset, and there exist strong mathematics proofs that demonstrate that it is robust to any post-processing of the released data. Differentially-private primitives, called \emph{mechanisms} have been designed for a wide range of operations~\cite{dwork2008differential} and are applied on different fields~\cite{xiong2014survey}, including social network analysis~\cite{jiang2021applications}, location dataset~\cite{kim2021survey} and healthcare~\cite{dankar2013practicing}. A more recent corpus of research focuses on designing differentially-private machine learning~\cite{ji2014differential} or deep learning~\cite{ha2019differential,el2022differential} algorithms.

Fewer works explored the use of differential privacy for network measurements. McSherry et al.~\cite{mcsherry2010differentially} initially propose the use of differential privacy to process network traces and report their experiences conducting a diverse set of analyses in a differentially private manner. Later, Zhoue et al.~\cite{zhou2015lightweight} propose a framework based on differential privacy to gather data in home networks. Focusing on browsing activity rather than on network traces, Fan et al.~\cite{fan2014monitoring} develop a system that extracts differentially private aggregates of web browsing histories to be released. With \TOOL, we complement the above works by providing a fully-fledged tool for differentially-private passive monitoring. To the best of our knowledge, we are the first to release an open-source tool with this goal.
\section{\TOOL}
\label{sec:metho}

\subsection{Objectives}

\begin{figure}
    \centering
    \includegraphics[width=1\columnwidth]{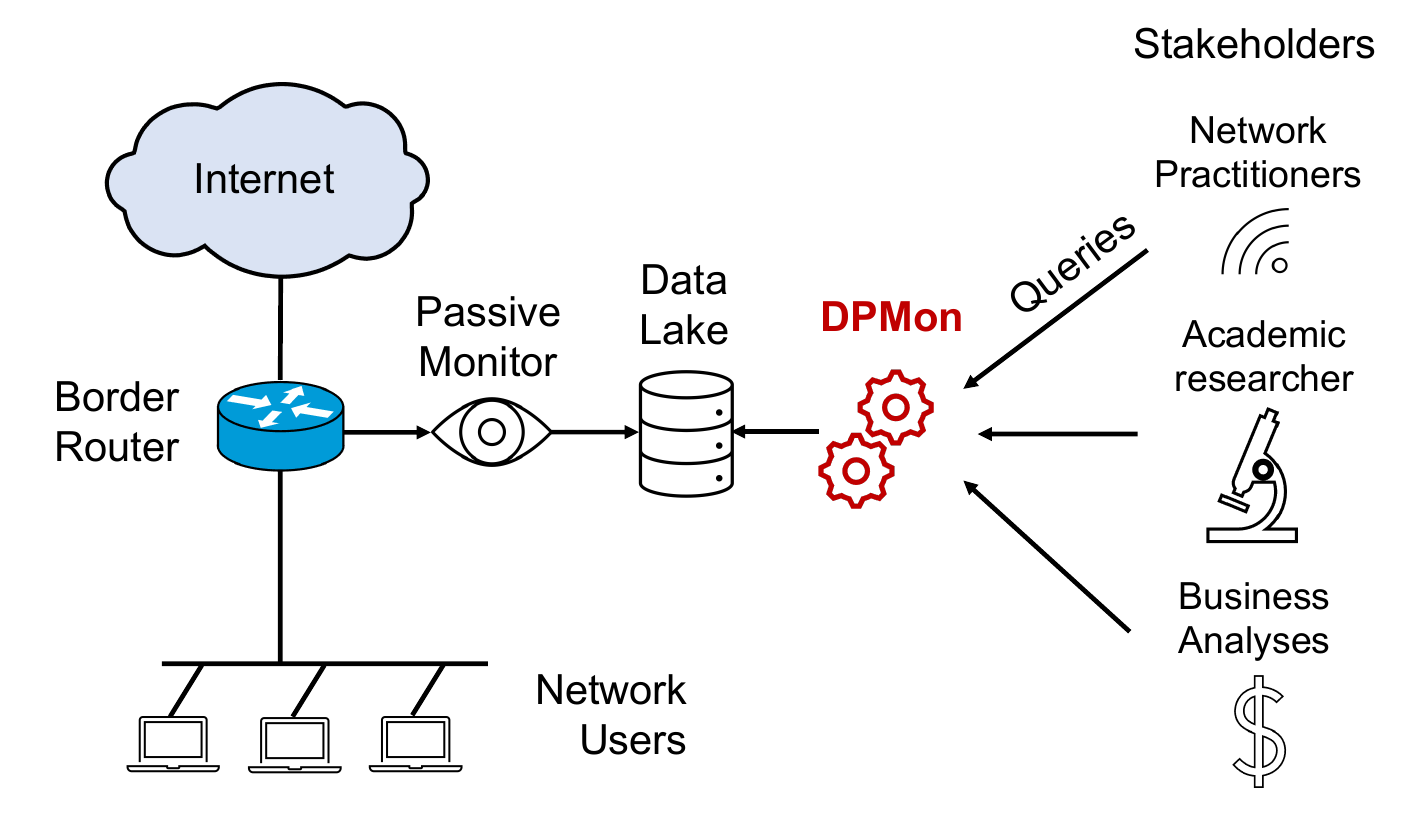}
    \caption{\TOOL typical deployment}
    \label{fig:dpmon}
\end{figure}

The goal of \TOOL is to run privacy-preserving queries to a dataset of passive network measurements. It applies Differential Privacy to the results of the query to control the contribution of a single individual to the query value. Thus, it mitigates the privacy risk of running and spreading network measurements, allowing network practitioners and stakeholders to safely share valuable information about network operation, Quality of Service and cyber security threats. In a typical operational deployment, sketched in Figure~\ref{fig:dpmon}, we suppose the network administrator operates a monitoring infrastructure composed of one or more network meters exporting measurements in the form of flow records. Indeed, \TOOL is designed to operate on flow records, a data format where each TCP or UDP flow constitutes an entry. Each flow is identified by its 5-tuple (IP addresses, port numbers and L4 protocol used) and described by a (potentially rich) set of features, such as packet number and size, domain name, and performance metrics (e.g., TCP Round-Trip Time). \TOOL operates as a query engine allowing stakeholders internal or external to the organization to gain insights on the network operation.

\subsection{Operation}
\label{sec:operation}

\begin{figure}
    \centering
    \includegraphics[width=1\columnwidth]{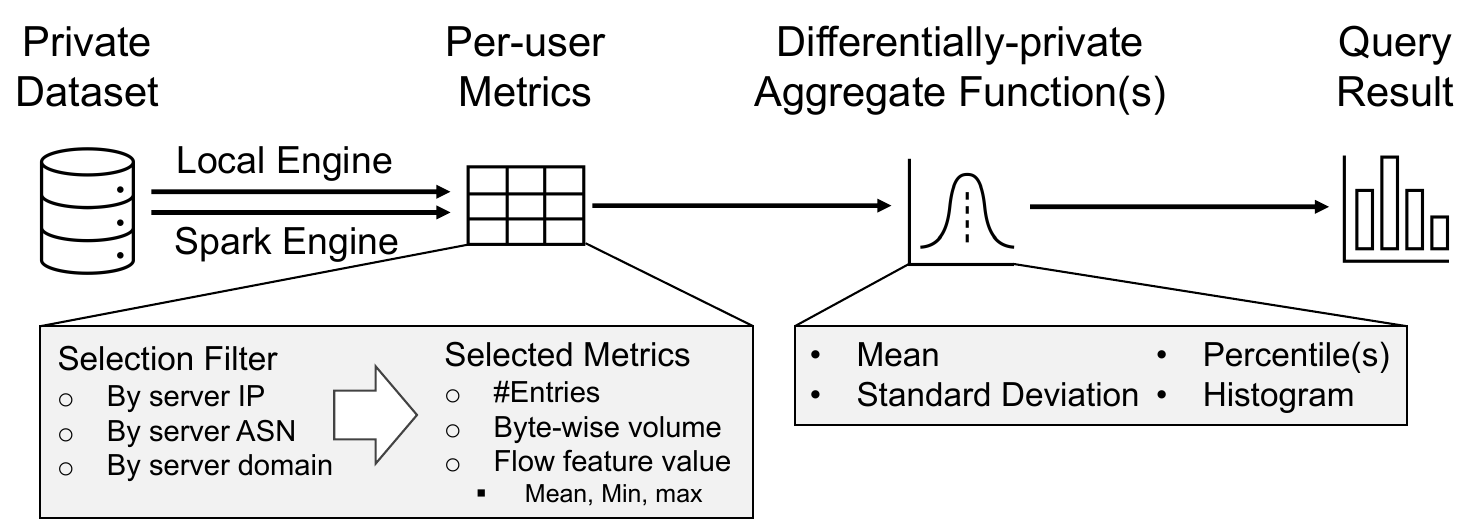}
    \caption{\TOOL query flow.}
    \label{fig:dpmon_flow}
\end{figure}

\TOOL operates as a query engine: It receives requests from the operator, processes themes and finally returns the results. The operator can choose among a set of predefined queries which can be configured to match the data of interest (e.g., flows to a specific server IP address). In all cases, the operator must specify which information shall be extracted from each flow and how to aggregate it at the per-user level. Then, per-user aggregates undergo a second round of aggregation using differentially-private aggregation functions and are finally returned to the operator as the final result.

In Figure~\ref{fig:dpmon_flow}, we sketch the whole data processing operation. \TOOL reads the private dataset or a partition of it. Typical queries involve a limited temporal span -- e.g., log files referring to a single day or week. Then, data are filtered according to the operator's preference, so that it is possible to isolate flows to specific entities. \TOOL can filter data by three criteria: (i) by server IP address; (ii) by server domain name, if the dataset includes this information,\footnote{Most flow exporters gather the server domain name by parsing the Server Name Indication Field of the TLS Client Hello packet or by inspecting DNS traffic.} (iii) by server Autonomous System Number. In the latter case, the system must be fed with an updated Routing Information Base (RIB) to associate an IP address to an Autonomous System Number. Notice that it is not possible to set a filter on the client IP address, as it is supposed to be private and hidden from the operator.

Out of the flows passing the given filter, \TOOL extracts the metric desired by the operator. At this stage, the system computes an aggregate for each user (i.e., client IP address), so that these aggregates can be later perturbed with differential privacy. Indeed, this step is fundamental as differential privacy protects individuals, thus it expects a record for each one. The operator can choose between the following per-user aggregates: (i) the number of entries (i.e., flows), (ii) the byte-wise downloaded, uploaded or total volume, (iii) the average, minimum or maximum value of a given flow feature.\footnote{Technically, aggregates different than average, minimum or maximum could be computed, but, practically, these are the most useful.} Notice that the latter feature depends on the dataset's nature. For example, the dataset could include a measure of the TCP round-trip time or the size of the first packets.

Then, \TOOL applies a differentially-private aggregation function to the per-user aggregates. To this end, it relies on the IBM DiffPrivLib library~\cite{holohan2019diffprivlib}, which implements the most common DP mechanisms. The operator configures the desired aggregate, which will be returned as the final result. The available choices are: (i) Mean, (ii) Standard Deviation, (iii) one (or more) percentiles, (iv) histogram.
For the first three options, a single number number is returned to the operator (two or more in cases it requests more than one percentile). If the operator requests a histogram, \TOOL computes the differentially private histogram of the per-user aggregates, which summarizes its distribution. This is the most powerful feature of \TOOL, as histograms carry a large informational value, while only a little noise must be added to achieve DP.

\begin{figure*}[t]
    \centering
    \includegraphics[width=\textwidth]{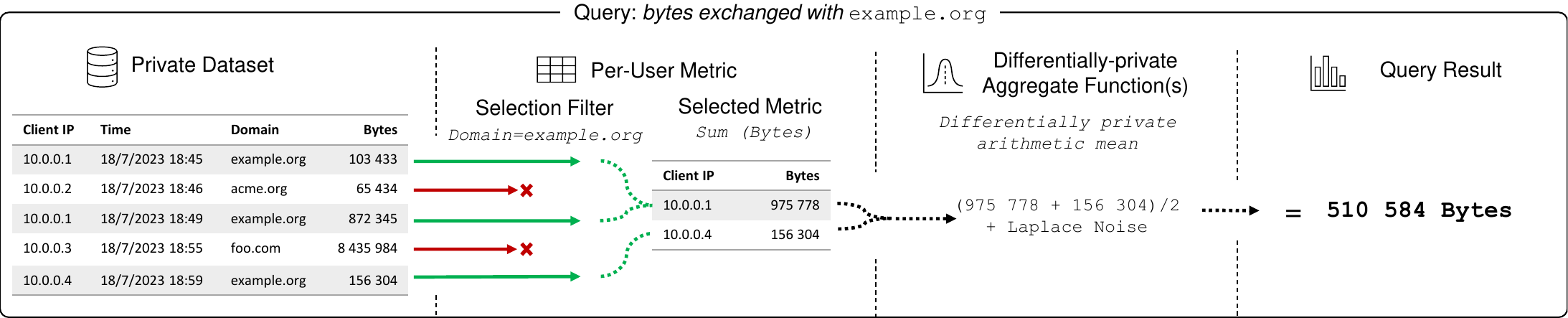}
    \caption{\TOOL query flow.}
    \label{fig:dpmon_example}
\end{figure*}

Figure~\ref{fig:dpmon_example} reports the steps in running an example query computing the average number of bytes a user exchanges with the \texttt{example.org} website. We assume \TOOL is operating on a dataset where each flow is annotated with the server domain name (e.g., extracted from the TLS handshake). First, the dataset is filtered according to the operator's criterion -- i.e., matching flows to the \texttt{example.org} domain. Second, \TOOL computes a per-user aggregate, in this case, the sum of the exchanged bytes. Third, a differentially-private aggregation function is applied to the set of per-user aggregates. In this example, \TOOL applies a differentially-private mean function, which adds a Laplace noise before releasing the output.

\paragraph{Operator's privacy budget}
The operator must specify the \emph{privacy budget} to allocate for each query. The privacy budget, called $\epsilon$ in the DP terminology, is a concept of differential privacy that represents the amount of information one can extract from the dataset. \TOOL allows to specify the privacy budget for each query, and, in a typical deployment, each \TOOL's operator holds a certain amount of privacy budget to spend. The operator can decide whether to run a few yet accurate queries or, conversely, many course-grained ones. The numeric amount of the privacy budget is not set by \TOOL, and the system administrator should impose a (potentially different) value for each operator. In the literature, it is still debated how to properly set $\epsilon$~\cite{hsu2014differential,naldi2015differential,dwork2019differential}, while values between 2 and 10 are commonly used in real systems~\cite{desfontaineslist}. Finding novel techniques to set $\epsilon$ is out of the scope of this work, although, in this paper, we show the impact of the DP privacy budget for network trace datasets.

\paragraph{Query Engines}
By default, \TOOL processes data locally using Python and the Numpy library. This is called the \emph{Local Engine}. As network measurements are typically large datasets and may account for several GB per day, \TOOL is designed to work above Apache Spark\footnote{\url{https://spark.apache.org/}} for scalable data processing. Thus, with \TOOL, it is possible to use the \emph{Spark Engine}, to leverage a big data cluster to process large quantities of data. We assume the \TOOL operator has a pre-configured Spark cluster in operation.

\paragraph{Supported Data Formats}
\TOOL must operate on flow records, where each flow describes a communication between two endpoints using the TCP or UDP protocols. Flow records always report endpoint identifiers (such as IPv4 or IPv6 address and port numbers) and minimal volumetric information (number of packets and bytes exchanged). Sophisticated flow exporters offer a number of additional metrics such as performance indicators or server domain names. \TOOL can virtually operate on any dataset of flow records, while, so far it supports two specific formats:
\begin{itemize}
    \item \textbf{NetFlow records}: as converted in \texttt{csv} format using the standard NFDump tool to convert a NetFlow database created by NFCapd.\footnote{These tools are available in most Linux distributions and online at \url{https://github.com/phaag/nfdump}.}
    \item \textbf{Tstat log files}: Tstat is a passive meter that exports rich flow records with hundreds of features, including packet and byte volume, TCP Round-Trip Time, etc.\footnote{\url{http://tstat.polito.it/}} Tstat extracts the domain name of the server from the TLS Server Name Indication header of from the HTTP \texttt{Host} header. \TOOL can benefit from the statistics provided by Tstat to offer insights on service usage and their performance. 
\end{itemize}

In general \TOOL could operate on any dataset containing entries that associate a user with some entity (a server IP address, a domain name, etc). That is, with minimal adaptation, it can process HTTP proxy or firewall log files, which typically contain data in this form.

\section{Deployment Use Case}
\label{sec:results}

In this section, we showcase \TOOL operation when used to monitor an operational network. We show that \TOOL provides useful insights which can safely be shared by the network operator with other parties. The extracted metrics help understand network traffic in terms of traffic volume, user habits and network performance. At the same time, \TOOL allows tuning the amount of information extracted from network logs, thus regulating the trade-off between data utility and users’ privacy.

\subsection{Monitoring Infrastructure and Dataset}

We deploy \TOOL on an existing monitoring infrastructure in a campus network. It consists of a passive probe, fed with the traffic leaving and entering the campus. To this end, we instrumented the border router of the organization to mirror all incoming and outgoing traffic to the passive probe through a ``span port''. The probe runs Tstat~\cite{trevisan2017traffic} to produce rich flow records reporting, for each TCP and UDP flow observed, several characteristics, including volumetric (number of packets and bytes transmitted) and performance metrics (TCP round-trip time, bitrate, etc.). Moreover, for each TLS and QUIC flow, Tstat records the fully qualified domain name as extracted from the Server Name Indication (SNI) field of the Client Hello messages.

We consider logs for a week in May 2023. The monitored audience includes approximately $20\,000$ students and $1\,000$ among professors and staff. For privacy reasons, the client IP of all flow is not captured and is replaced with an anonymized identifier. In total, we collect $400M$ flow records accounting for $135$ GB of compressed log files. The traffic includes mostly HTTPS and QUIC sessions to popular websites, portals and online social networks. The data are stored on a medium-sized big data cluster using Ceph as distributed storage. We use \TOOL to query the data using the Apache Spark engine running over a Kubernetes cluster.



\subsection{The role of the privacy budget $\epsilon$}

\begin{figure}
    \centering
    \begin{subfigure}[t]{0.23\textwidth}
        \includegraphics[width=\columnwidth]{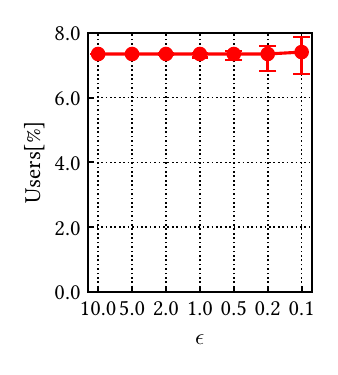}
        \vspace{-7mm}
        \caption{\texttt{wikipedia.org}}
        \label{fig:eps1}
    \end{subfigure}
    \hspace{-5mm}
    \begin{subfigure}[t]{0.23\textwidth}
        \includegraphics[width=\columnwidth]{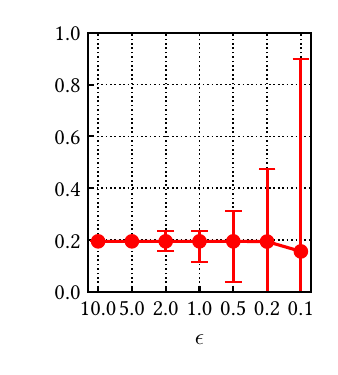}
        \vspace{-7mm}
        \caption{\texttt{physicsforums.com}}
        \label{fig:eps2}
    \end{subfigure}
    \caption{Distribution of the results of a query computing the share of users accessing a given website with different privacy budgets $\epsilon$. The marker represents the median, the bars span from the $5^{th}$ to the $95^{th}$ percentile.}
    \label{fig:eps}
\end{figure}

We first illustrate the role of the privacy budget $\epsilon$ on the results of the queries. As the privacy budget indirectly controls the amount of information to be extracted from the data, the smaller the budget, the larger the noise added to the query result. We show its impact on a simple query that computes the share of users accessing a given website. Here, a second aspect of DP plays a crucial role. It protects individuals, thus the noise added using a differentially-private aggregate function decreases with the number of users involved in the aggregation. Specifically, this query works by computing a two-bin histogram, representing the number of users i) \emph{accessing} and ii) \emph{not accessing} the given website. The amount of noise added to a small bin is thus (relatively) much higher than the one added to large bins. This aspect entails that querying for a rare website (accessed by a small fraction of users) returns a much less accurate measure than querying a popular one. This is an expected and desired effect, as the information that a user visits a rare website is far more informative (and privacy sensitive) than knowing that she accessed a popular portal, thus it must be protected to a larger extent. 

Figure~\ref{fig:eps} shows the result of the query, separately for two websites: (i) a popular one, \texttt{wikipedia.org}, and (ii) a niche one, \texttt{physicsforum.com}. The $x$-axis represents the query result with different $\epsilon$, while the $y$ reports the share of users accessing the website. Each query is repeated 100 times, and the figure reports the median, $5^{th}$ and $95^{th}$ percentiles. Starting from \texttt{wikipedia.org} (Figure~\ref{fig:eps1}), we observe that almost 8\% of users acceded the website in the given period. With $\epsilon \geq 1$, the measurement error is negligible. Indeed, \TOOL applies DP to histograms where both bins include thousands of individuals, making the resulting perturbation relatively small. With smaller budgets, the measurement comes with increased variability, but even with $\epsilon=0.1$ it
does not vary by more than 10\%.

Different is the case for an uncommon website such as \texttt{physicsforum.com}. We obtain a stable estimation with $\epsilon \geq 5$, revealing that only $\approx 0.2\%$ of users access it (equivalent to a few dozen individuals). With small budgets, the variability explodes, showing that, when properly set, DP protects the privacy of users, preventing infrequent attributes from being disclosed.

In summary, these figures exemplify how the privacy budget $\epsilon$ plays a crucial role in tuning the tradeoff between data utility and privacy. However, as discussed in Section~\ref{sec:operation}, there is no globally-accept range of values for $\epsilon$, that should be tuned on the specific context. In this work, we do not formulate such recommendations, We limit ourselves to observing how $\epsilon$ greater than 5 allows measuring with good accuracy the number of users accessing even unpopular websites, which might not always be desirable.

\subsection{Measuring distributions}

One of the most powerful features of \TOOL is its ability to compute differentially private histograms. This allows for studying the distribution of a given metric across users, which has a more informative value than a mere statistical index (such as mean or standard deviation). \TOOL computes histograms by computing a per-user metric and then counting how many of them fall in different bins. The size and number of bins can be automatically computed by \TOOL or can be specified by the operator. The latter possibility allows the use of non-standard bins in case it is known the distribution has a very skewed shape and requires e.g., logarithmic bins.

\begin{figure}[t]
    \centering
    \includegraphics[width=0.45\textwidth]{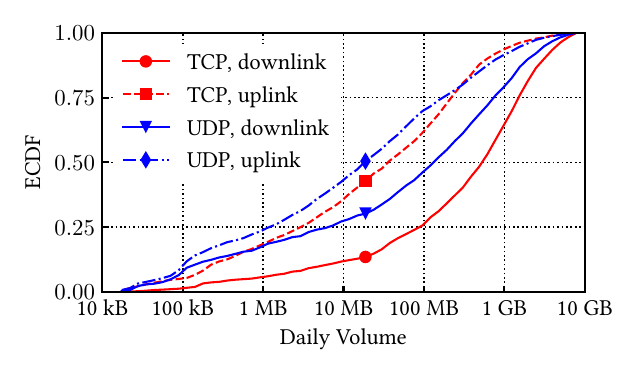}
    \caption{Distribution of the weekly per-user volume, separately by direction and Layer-4 protocol. We set $\epsilon=1$.}
    \label{fig:volume}
\end{figure}

We showcase this feature with Figure~\ref{fig:volume}, which shows the Empirical Cumulative Distribution Function of the per-user volume computed using the data extracted from histograms provided by \TOOL. Separately for traffic direction (incoming or outgoing) and by Layer-4 protocol (TCP or UDP), we compute the histogram of the weekly per-user volume. As the selected metric spans different orders of magnitude (from a few MB to several GB), we set a custom list of 100 bins arranged on a logarithmic scale -- i.e., bins for small values are smaller. We make four queries to cover the four combinations of direction and Layer-4 protocol, each using a privacy budget $\epsilon=0.25$. We discard bins referring to users generating less than 10~kB, that we consider inactive.

The results in Figure~\ref{fig:volume} provide several takeaways. First, in median users generate less than 1~GB of traffic in downlink and few tens of 10MB in uplink. TCP is still prevalent over UDP, likely because the traffic is from an educational institution rather than residential subscriptions. Notice that the figure has been computed consuming only a total privacy budget of $1$ (four queries with $\epsilon=0.25$). Indeed, when computing distributions, DP is well suited as it easily operates on histograms. If histogram bins are properly chosen to include a non-negligible number of individuals, the amount of noise added to each does not prevent the observer from obtaining faithful insights.

\section{Limitations and Future Developments}

This paper introduced \TOOL, a tool for executing privacy-preserving queries to a passive monitoring infrastructure. To ensure the anonymity of query results, we have opted to utilize DP, enabling the perturbation of query outputs to restrict the contribution of any single individual to the final results below a controlled threshold. \TOOL facilitates different stakeholders in gaining insights into network activity, as evidenced by our deployment use cases, which demonstrate the extraction of classical network monitoring figures while simultaneously upholding users' privacy.

Currently, \TOOL operates as a programming library intended for use by network administrators for data extraction and is not intended for direct use by external analysts. Consequently, network administrators should consider developing a web service atop \TOOL to facilitate external usage and enforce relevant policies and privileges (which depend on the specific use case). Notably, \TOOL has already been employed in a research study involving passive measurements~\cite{perdices2022satellite}, establishing it as our preferred method for processing such data in future research endeavours.

While \TOOL has DP at its core and employs various techniques to compute differentially-private aggregates, these techniques may impose limitations on the types of queries feasible. Specifically, while DP effectively protects users and calculates the contribution of individual users to query results, it can be restrictive in certain cases. Currently, direct exportation of (anonymized) flow records or traffic samples using DP is not feasible. Additionally, due to the nature of DP, all queries must initially aggregate the desired metric at the per-user level before applying a differentially-private noisy function. In certain scenarios, this mandatory aggregation step may be undesirable, necessitating the design of novel algorithms tailored to address this specific use case.

\section*{Acknowledgments}
\noindent
This work has been supported by the GÉANT Innovation Programme with the project ``Differential Privacy for Networks'' and by the Italian Ministerial Grant PRIN 2022 with the project ``COMPACT: Compressed features and representations for network traffic analysis in centralized and edge Internet architectures'' (grant no. 2022M2Z728). This manuscript reflects only the authors' views and opinions.

\bibliographystyle{plain}
\bibliography{biblio}

\end{document}